\documentstyle[12pt]{article}
\begin{document}
\begin{center}
\vspace{1.5in}
{\LARGE

Structure of A=6 Nuclei: ${^6}He$, ${^6}Li$ and ${^6}Be$ }

\end{center}
\vspace{.4in}
\begin{center}
{\bf Afsar Abbas}\\
\vspace{.1in}
Institute of Physics\\ 
Bhubaneshwar-751005, India\\
email: afsar@iopb.res.in
\end{center}
\vspace{1.2in}
\begin{center}
{\bf Abstract}
\end{center}
\vspace{.3in}

It is commonly believed that ($\alpha$-d) and (${^3}He$ - ${^3}H$) 
represent equivalent states of ${^6}Li$. It is shown here that this is not 
correct. These two are actually orthogonal to each other. It is shown here
that these two with very different shapes and forms actually co-exist for 
the ground state of ${^6}Li$. This shape co-existence is the same as similar
phenomenon in heavy nuclei. The puzzling anomaly of extremely small 
branching ratio for beta delayed deuteron emission in ${^6}He$ is 
explained here. In addition the anomalously large branching ratio for beta 
delayed triton emission in ${^8}He$ is explained. The cluster structure of 
the ground state and of the low-lying states of ${^6}He$, ${^6}Li$ and 
${^6}Be$ is clarified. 

\newpage

Cluster structure forms an important aspect of modern nuclear studies 
[1,2]. In addition, unexpected halo structure has been found to exist in 
neutron-rich light nuclei [3,4]. Here we wish to study A=6 nuclei: 
${^6}He$, ${^6}Li$ and ${^6}Be$. 
The aim is to study as to what kind of clusters are relevant here 
and how these affect the structure of these nuclei in the ground state and 
in the excited states. Recent publication [5] of the energy levels of 
these nuclei shall be found to be handy for our studies here. 

It has been discussed extensively how ${^6}He$ is a two-neutron halo 
nucleus.  In this case two neutrons are found to be very loosely bound around 
an inert $\alpha$-particle [3]. This has been well studied and therefore one 
can safely say that the predominant structure of the ground state of 
${^6}He$ is ($\alpha$-n-n). 
The other possibility of (t-t) cluster structure does not seem to have a 
role in the ground state of ${^6}He$, but should have a role specially for the
excited states above the (t-t) breakup threshold.  But why (h-h) cluster
structure does not mix with ($\alpha$-n-n) structure for the ground state of
${^6}He$ is not understood so far. Below we shall give an explanation of this
effect (and also for the corresponding situation in ${^6}Be)$. ${^6}Be$ being
mirror partner of ${^6}He$ would be expected to have similar structure -
ie. predominant ($\alpha$-p-p) for the ground state. This similarity seems to 
hold well. But the same cannot be said for the isospin partner ${^6}Li$. 

It is generally believed that for the ground state and the low excited
states of ${^6}Li$ ($\alpha$-n-p) (or ($\alpha$-d)) configuration is much more
important than (h-t) (note: h=${^3}He$) [eg. 6 (p.115), 7 (p.104)]. Most of 
the studies recently follow this line of ($\alpha$-n-p) being significant for 
the ground state and the low-lying excitations [8,9]. Meanwhile there have 
been compelling experimental evidences which indicate that the (h-t) 
configuration for ${^6}Li$ cannot be ignored, and may be quite important 
[10-12]. It turns out that the two cluster configurations ($\alpha$-d) and 
(h-t) are not mutually exclusive and that they can both be present as they are 
not orthogonal either [6,13]. A few recent theoretical studies also support 
this view [14]. Recent experimental study of tri-nucleon cluster knockout 
for ${^6}Li$ favours simultaneously existing ($\alpha$-d) and (h-t) structures 
[15]. Most of the scientists are by and large reconciled to the situation as 
depicted in the following quotation from ref. 15 : "Since the alpha particle 
is a very tight system, one would expect ${^6}Li$ to be most of the time in an 
($\alpha$-d) configuration (even though the (${^3}H$ - ${^3}He$) configuration
occurs some of the time)". The vagueness conveyed by this statement can be 
traced to the fact that ($\alpha$-d) and (h-t) configuration of 
${^6}Li$ are not orthogonal and it is hard to give physical interpretation 
to the relative weights of these configurations [6,p.43]. Let us look into 
this equivalence, simultaneous existence and non-orthogonality of 
($\alpha$-d) and (h-t) cluster structures of ${^6}Li$. 

Before we do this, let us point out one further information on $^{6}Li$, 
which is anathema for all other models. This is the study of beta-delayed 
deuteron emission in ${^6}He$ and beta-delayed triton emission in ${^8}He$ 
[16, 17]. The most established beta decay of ${^6}He$ 
is to the $1^+$ ground state of ${^6}Li$. They find [16,17] evidence of a
second beta decay branch leading to unbound final state consisting of a
deuteron and an alpha particle. They find "... the astonishingly small
branching ratio for deuteron in ${^6}He$ and the surprisingly large
one for tritons in ${^8}He$ " [17]. This one order to two order suppression 
cannot be understood on the basis of any models of ${^6}Li$.  
In this letter we give an explanation of this anomaly in terms of the
model that we shall develop here. This model will enable us to have a
better and more consistent understanding of these three A=6 nuclei. 
In addition the anomanously large triton branch case
for ${^6}He$ will also thus be naturally explained.

In oscillator model the ground state of ${^6}Li$ is described by the 
following wave functions of equivalent forms [6,p.40, appendix C] 

\begin{equation}
 \psi_{gs} (^{6}Li) = A \{ \phi_{0}(\alpha) \phi_{0}(d) 
                        \chi(\alpha - d) \}
 = N A ( \phi_{0} (t) \phi_{0} (h) \chi(t-h) \}
\end{equation}

Here $\chi (\alpha-d)$ and $\chi (t-h)$ are two oscillator quanta wave
functions and 
$\phi_{0} (\alpha), \phi_{0} (d), \phi_{0} (t), \phi_{0} (h) $
are the internal wave functions. The constant N
is here to ensure that the two functions are equal to each other. The
mathematical equivalence of the two can be easily demonstrated.  Since
they are the same, these can exist simultaneously 
and also note that these are not
orthogonal [6]. None of this is changed even with the hard core of the
Jastrow form in the two body interaction [6, p.113]. This is the broad 
view which has been dominating the ${^6}Li$ studies so far [11-15]. 

However, in the above mathematical proof, there is an underlying ansatz 
- which is that in the Schroedinger equation one need go only up to 
two-body interactions and ignore all many body forces like the Three Body 
Forces (3BF) $V{_{ijk}}$ [6, p.1]. This was perhaps acceptable in the 60's 
but today we know that for A=3 nuclei 'h' and 't' there is no 
way we can obtain the binding energies and the rms radii for 
these nuclei without invoking some form of a simple or a more 
complicated 3BF [eg. references in 18,19] in addition to the 
usual two body interactions. In addition in A=4 nuclei Four Body Forces
( 4BF ) may be active too. Once the 3BF ( and 4BF ) are included, clearly 
the equivalence (and simultaneous existence) 
of ($\alpha$-d) and (h-t) clusters for ${^6}Li$ is 
lost. Hence all the physics which has been obtained with this idea 
[6, 11-15, etc] cannot be accepted any more. 

Having disposed of the equivalence and simultaneous existence of 
($\alpha$-d) and (h-t) 
clusters for ${^6}Li$, could it be that they are still not orthogonal? 
In this connection it may be pointed out that while discussing the coupled 
channel equations for nuclear reactions and nuclear bound states, 
Wildermuth and McClure [6,p.88] do require that these two configurations 
be "asymptotically orthogonal".

To discuss this non-orthogonality further we have to understand some of
the present author's recent work on halo nuclei [20]. Basing arguments on QCD 
and quark model [21,22] one is able to explain the existence of hole in 
central charge density distribution of A=3 and A=4 nuclei, the existence
of haloes in neutron-rich nuclei, the occurrence of nuclear molecules and 
the onset of clustering in nuclei [20]. Interestingly, it is QCD and 
quark-based considerations which are able to explain these low-energy and 
ground-state properties of nuclei. 

The central empirical fact (which arises from QCD and quark model basis) 
is the unique charge and matter distribution of A=3 and A=4 nuclei. That
is the "hole" in these nuclei [21,22]. It is well known that 
in the nuclei 'h', 't' and $\alpha$ 
the central densities are larger than any other nuclei. In addition, the 
fact that there is a hole at the centre actually pushes the densities 
outwards and making them peak further out in these nuclei. This is a unique 
property of these nuclei [20-22]. In addition, deuteron is very large and 
diffuse. Thus, ($\alpha$-d) and (h-t) configurations would appear very
different physically. It would not be trivial to change from one density 
distribution to another. This will be resisted by the basic difference in 
the two configurations. Hence it is on this physical ground that these two 
clusters : ($\alpha$-d) and (h-t) should be orthogonal to each other. 

Note that 'n' and 'p' are members of an isospin doublet. These combine 
together to give a bound triplet state (S=1, T=0), that is deuteron with a 
binding energy of 2.2 MeV. It has no excited states. The singlet state 
(S=0, T=1) is unbound by 64 keV. This being isospin partner of (n-n) (S=0, 
T=1) and (p-p) (S=0,T=1), all these are unbound. 
Now it turns out that 'h' and 't' are also members of a good isospin 1/2. 
Note that though both 'n' and 'p' are composites of three quarks 
these still act as elementary particles as far as low-energy 
excitations of nuclear physics are concerned. Only at relatively higher 
energies does the compositeness of 'n' and 'p' manifest itself. Similarly
the binding energies of ${^3}He$ and ${^3}H$ are 7.72 MeV and 8.48 MeV 
respectively and also these two have no excited states. Hence for low-energy 
excitations, of a few MeV, we may consider these as elementary. Their 
compositeness would be manifested at higher excitation energies. That is 
we treat 'h' and 't' as elementary isospin 1/2 entities here. This is 
similar to the two nucleon case. Hence, we would expect for (h-t) the 
triplet (S=1,T=0) to be bound and singlet (S=0,T=1) to be unbound. Also, its 
isospin partners (h-h) (S=0, T=1) and (t-t) (S=0, T=1) would be unbound too. 

This means that (h-h) (S=0, T=1) being unbound would not mix with the 
bound ($\alpha$-n-n) configuration of ${^6}He$ for the ground state. 
Similarly, the (t-t) (S=0, T=1) being unbound would not mix with 
the bound ($\alpha$-p-p) ground state configuration of ${^6}Be$. 
Therefore, ${^6}He$ and ${^6}Be$ in its ground 
state and low excited states remain pure ($\alpha$-N-N) configuration. 
This is the configuration that others have been discussing and using 
for these nuclei anyway [8,9]. However, things are different for ${^6}Li$. 

Here in ${^6}Li$ the (h-t) (S=1, T=0) configuration is 
bound with no other excited state. Being bound, it is close to the bound 
ground state of the orthogonal configuration ($\alpha$-d). Thus for the
ground state of ${^6}Li$ the two states (h-t) (S=1, T=0) and 
($\alpha$-d) (S=1,T=0) which though orthogonal actually co-exist. These 
being of such different physical shapes ( as we discussed before ) thus this 
should actually be viewed as co-existence of different shapes. This situation 
is similar to the co-existence of different shapes - spherical and deformed, 
etc. as is well known in the case of heavy nuclei. Note that here we are 
finding co-existence of two different shapes in a very light nucleus. 

Thus we expect that for the ground state of ${^6}Li$ (S=1, T=0), these two 
different cluster structures ($\alpha$-d) and (h-t) would mix and
repel. Thus, for the ground state wave function would be 

\begin{equation}
 \psi (^{6}Li) ) =  a  \psi (\alpha -p-n) + b \psi (h-t) 
\end{equation}

with $a^2 + b^2 = 1$. The amplitude 'a' and 'b' would be found from best 
fits to suitable experiments. As per various studies [8,9], 
one would expect 'a' to be more
dominant for the ground state, but that 'b' is not negligible [12,14] and 
actually should be quite significant [10,11,15]. 

If this be so, there should exist another higher-lying (S=1,T=0) 
orthogonal state in ${^6}Li$ with wave function 

\begin{equation}
 \psi ^{*} (^{6}Li) ) =  b  \psi (\alpha -p-n) - a \psi (h-t) 
\end{equation}

with 'a','b' the same as in $\psi(^{6}Li)$. Indeed, this state exists as 
the 5.65 MeV excited state in ${^6}Li$ [5] 
and we identify $\psi ^{*} (^{6}Li)$ with this. So both the ground state 
and the 5.65 MeV states are mixed states of these two orthogonal cluster 
structures. Hence, rest of the low-lying states in ${^6}Li$ 
(as in ${^6}He$ and ${^6}Be$) should be of simple ($\alpha$-N-N) structure. 

Note that for $\psi (\alpha$-p-n) amplitude 'a' is much larger than the 
$\psi$ (h-t) amplitude for the ground state and thus the $\psi (\alpha$-p-n) 
or $\psi (\alpha$-d) configuration would be much smaller than the 
corresponding $\psi$ (h-t) amplitude for the 5.65 MeV (1+,0) state. 
This will help us to explain the beta-delayed deuteron emission 
puzzle referred to earlier [16,17]. 

The normal beta decay of ${^6}He$ proceeds almost entirely to the $1^+$ 
ground state of ${^6}Li$ which as discussed is mainly of ($\alpha$-d) 
configuration. The other beta-delayed deuteron emission in our picture goes 
almost entirely to the 5.65 MeV $1^+$ state in  ${^6}Li$. Therefore in our 
model the weakness in the intensity of the branch is mainly due to the small 
fraction of ($\alpha$-d) existing in that state. 

The authors of ref. [16] had tried to bring in this $1^+$ excited state at 
5.65 MeV, but failed to reach the right conclusion mainly because firstly 
they also tried to simultaneously include the ground state in their
analysis, and secondly as they believed that the 5.65 MeV is mainly 
made up of ($\alpha$-d) configuration in the d-wave. The right answer is 
that this second branch proceeds only to the 5.65 MeV 
state which is mainly of (h-t) configuration as shown above. 

In trying to understand the surprisingly large branching ratio for 
beta-delayed triton emission in ${^8}He$, the authors of ref. [17] suggeted 
the existence of a new $1^+$ state at 9.3 MeV in ${^8}Li$. They did not 
know the origin of this state though. Quite clearly in our model 
here this is the state built up on the $1^+$ state at 5.65 MeV in ${^6}Li$.
Thus this excited $1^+$ state in ${^8}Li$ has two components which both 
have quantum numbers of the cluster structure ($\alpha$-t-n). Therefore 
tritons would be easily emitted in this branch of beta decay in ${^8}He$.
Thus this puzzle finds natural explanation in our model.

Another confusing situation which finds simple clarification in our model
is that of (${^6}Li$,d) and (${^7}Li$,t) reactions. On the basis of the 
misconception that only ($\alpha$-d) and ($\alpha$-t) give the cluster 
structures of ${^6}Li$ and ${^7}Li$ respectively, it is commonly
suggested that the the above two reactions should proceed through
the $\alpha$-particle stripping, similar to the (d-p) reaction.
But careful calculations by various groups have shown that this is more
complicated than the simple neutron stripping case. Hence these two 
reactions are much more than a single $\alpha$ stripping. In our model 
this is simply understood as arising from the two components of the ground 
state wave function of ${^6}Li$ and ${^7}Li$.

In summary, it is shown here that contrary to all earlier expectations, 
the (h-t) and ($\alpha$-d) structures are orthogonal to each other in
${^6}Li$. Therefore, for the ground state of ${^6}Li$, these two different 
shapes co-exist. We have shown that, for ${^6}He$ and ${^6}Be$ the 
configuration for the ground state and low excitations is mainly 
($\alpha$-N-N). In ${^6}Li$ it is ($\alpha$-N-N) for all low-lying 
states except the ground state and the 5.65 MeV state ( both $1^+$,0) 
which are mixed states of the orthogonal clusters ($\alpha$-d) and (h-t). 
Thus, in this model one is able to explain the puzzling extremely small 
branching ratio of beta-delayed deuteron emission in ${^6}He$. Simple 
extrapolation of these states to ${^8}Li$ explains the equally puzzling 
large branching ratio of beta-delated triton emission in ${^8}He$.

\vspace{.8in}

{\bf References} 

\vspace{.2in}

1 P.E. Hodgson, Contemp. Phys. {\bf 43} (2002) 461. 

\vspace{.1in}

2. B.R. Fulton, Contemp. Phys. {\bf 40} (1999) 299.

\vspace{.1in}

3. I.Tanihata, D. Hirata, T. Kobayashi, S. Shimoura, K. Sugimoto and
H. Toki, Phys. Lett. {\bf B289} (1992) 261

\vspace{.1in}

4. J.S. Al-Khalili and J.A. Tostevin, Phys. Rev. Lett.. {\bf 76} 
(1996) 3903.

\vspace{.1in}

5. D.R. Tilley, C.M. Cheves, J.L. Godwin, G.M. Hale, H.M. Hofmann, 
J.H. Kelly, C.G. Sheu \& H.R. Weller, Nucl. Phys. 
{\bf A708} (2002) 3.

\vspace{.1in}

6. K. Wildermuth and W. McClure, Springer Tracts in Modern Physics, 
{\bf 41} (1966),  Springer Verlag, Berlin. 

\vspace{.1in}

7. K. Langanke, Adv, Nucl. Phys. {\bf 21} (1994) 82.

\vspace{.1in}

8. M.U. Zhukov, B.V. Danilin, A.A. Korsheninnikov, L.V. Chulkov,
Nucl. Phys. {\bf A538} (1992) 375c.

\vspace{.1in}

9. V.I. Kukulin, V.N. Pomerantsev, Kh.D. Razikov, V.T. Voronchev 
and  G.G. Ryzhick, Nucl. Phys. {\bf A586} (1995) 151.

\vspace{.1in}

10. F. Young, P.D. Forsyth and J.D. Marion. Nucl. Phys. 
{\bf 91}(1971) 209.

\vspace{.1in}

11. M.F. Werby, M.B. Greenfield, K.W. Kemper, D.L. McShan
and S. Edwards, Phys.Rev. {\bf C8} (1973) 106.

\vspace{.1in}

12. R.G. Lovas, A.T. Kruppa, R. Beck and F. Dickmann,
Nucl. Phys. {\bf A474} (1987) 451.

\vspace{.1in}

13. R.D. Amado and J.V. Noble, Phys. Rev. {\bf C3} (1971) 2494.

\vspace{.1in}

14. A.Csoto, Phys, Rev. {\bf C49} (1994) 3035.

\vspace{.1in}

15. J.P. Connelly, B.L. Berman, W.J. Briscoe, K.S.Dhuga, A. Mokhtari, 
D. Zubanov, H.P. Blok, R. Ent, J.H. Mitchell and L.Lapikas, 
Phys. Rev. {\bf C57} (1998) 1569.

\vspace{.1in}
	
16. K. Riisager, M.J.G. Borge, H. Gabelmann, P.G. Hansen, L. Johannsen, 
B. Jonson, W. Kurcewicz, G. Nyman, A. Richter, O. Tengblad
and K. Wilhelmsen,  Phys. Lett.. {\bf B235} (1990) 30.

\vspace{.1in}

17. M.J.G. Borge, L. Johannsen, B. Jonson, T. Nilsson, G. Nyman, 
K. Riisager, O. Tengblad and K. Wilhelmsen Rolander, 
Nucl. Phys. {\bf A560} (1993) 664.

\vspace{.1in}

18. Proc. Few Body Problem in Physics Conf., Taiwan 2000, 
Nucl. Phys. {\bf A684} (2001).

\vspace{.1in}

19. Proc. Few Body Problem in Physics Conf., Portgual 2000, 
Nucl. Phys. {\bf A689} (2001) Nos. 1, 2.

\vspace{.1in}

20. A. Abbas, Mod. Phys. Lett. {\bf A16} (2001) 755.

\vspace{.1in}

21. A. Abbas, Phys. Lett. {\bf B167} (1986) 150. 

\vspace{.1in}

22. A. Abbas, Prog. Part. Nucl. Phys. {\bf 20} (1988) 181.

\end{document}